\pdfoutput=1
\documentclass[12pt,titlepage]{article}

\font\grande=cmr9.5 scaled \magstep4
\font\medio=cmr9.5 scaled \magstep2
\outer\def\beginsection#1\par{\medbreak\bigskip
      \message{#1}\leftline{\bf#1}\nobreak\medskip
\vskip-\parskip
      \noindent}
\usepackage{graphicx} 
\usepackage{mathrsfs}
\begin{document}

\bibliographystyle {unsrt}

\titlepage

\vspace{1cm}
\begin{center}
{\grande Planckian hypersurfaces, inflation and bounces}\\
\vspace{1cm}
Massimo Giovannini \footnote{e-mail address: massimo.giovannini@cern.ch}\\
\vspace{1cm}
{{\sl Department of Physics, CERN, 1211 Geneva 23, Switzerland }}\\
\vspace{0.5cm}
{{\sl INFN, Section of Milan-Bicocca, 20126 Milan, Italy}}
\vspace*{1cm}
\end{center}
\vskip 0.3cm
\centerline{\medio  Abstract}
\vskip 0.1cm
When the different wavelengths of the scalar and tensor modes of the geometry are all 
assigned on the same space-like hypersurface the maximally amplified frequencies of the spectrum 
remain smaller than the Planck mass only if the duration of a stage of accelerated expansion and the corresponding 
tensor to scalar ratio are severely constrained. All the different wavelengths can be initialized on the same space-like hypersurface  at the onset of inflation but this strategy and the related conclusions are plausible only
for classical inhomogeneities. We argue that a whole class of potential constraints is easily evaded provided the 
different wavelengths of the quantum fields are assigned as soon as they cross the corresponding 
Planckian hypersurfaces. In this case the Cauchy data for the mode functions depend on the wavenumber so that  larger wavelengths start evolving earlier while shorter wavelengths are assigned later.  Within this strategy the duration of a conventional inflationary phase and the corresponding tensor to scalar ratio are not constrained but the large-scale power spectra inherit specific large-scale corrections that remain however unobservable. 
\noindent
\vspace{5mm}
\vfill
\newpage
Already after the discovery of black hole evaporation \cite{haw1,haw2} it was noted  
 that the frequencies of massless species along the paths emerging from the past 
 null infinity and heading towards ${\mathscr I}^{+}$ can experience arbitrarily large 
 redshifts as the particles pass though the collapsing dust cloud prior to the formation 
 of the event horizon. The range of frequencies that can be seen by distant observers at late times 
would have had to originate at ${\mathscr I}^{-}$ with ultrahigh frequencies including frequencies 
above the Planck scale. Local Lorentz invariance would be violated if such frequencies were 
arbitrarily cut-off. Many questions have been raised through the years concerning the 
survival of Hawking radiation in the case of a breaking of local Lorentz-invariance (see e.g. 
\cite{jac1,unr1,deser1,deser2,park1}). With the purpose of investigating the stability of the 
process of black hole evaporation the dispersion relations have been purposely modified above a 
(nearly Planckian) energy scale. Different physical approaches led independently to the 
conclusion that the thermal emission is likely not to be destroyed by quantum gravitational 
effect even if  the tools employed to deduce the black hole evaporation may not be valid for 
arbitrarily small wavelengths \cite{unr1,deser1,park1}. 

The very same objection raised in the case of Hawking radiation has been subsequently 
brought up in the discussion of the inflationary power spectra since arbitrarily small 
wavelengths may appear when a phase of accelerated expansion is about to start 
\cite{kem,gree,kan1,staro1,bra,dan,mgio1,mgio2,action6,action9}. All along the past 
two decades these effects have been analyzed \cite{action10,action11a,action11b,action12,action13,action14} 
and there is consensus that physical frequencies larger than a certain reference energy (be it for instance
 the Planck or string mass) could mildly modify the scalar and tensor power spectra at large scales. 
 However these modifications have not been observed in spite of the repeated observational scrutiny 
 \cite{action13,action14,action14a} so that it is fair to say (as in the case of black hole evaporation) 
 that  the discussion of the scalar and tensor power spectra is not crucially affected by the wavelengths shorter 
 than the Planck length even if the tools used for the actual derivation could well be invalid for arbitrarily short wavelengths. 
 In spite of the latter conclusion it has been recently suggested \cite{FFF1,FFF2} (see also \cite{FFF3,FFF4}) that frequencies 
 above the Planck energy scale forbid, in practice, any sufficiently long conventional stage of inflationary expansion 
 and imply anyway severe bounds on the tensor to scalar ratio (e.g. $r_{T} < {\mathcal O}(10^{-30})$). According to this 
 further reprise of the original theme, if relic gravitons will ever be observed through a suitable $B$-mode 
 polarization \cite{FFF5} they will not come from conventional inflationary scenarios since the techniques 
used to derive the scalar and tensor power spectra are not consistent.

To spell out more clearly the terms of the problem it is useful to remind that in a conformally flat background of Friedmann-Robertson-Walker type the physical (i.e. $\lambda_{\mathrm{ph}}$) and the comoving (i.e. $\lambda$) wavelengths of the scalar and tensor modes of the geometry are notoriously related as $\lambda_{\mathrm{ph}}(\lambda, \tau) = \lambda \, a(\tau)$, where $a(\tau)$ is the scale factor and $\tau$ is the conformal time coordinate. As $\tau \to - \infty$ there are two extreme physical possibilities: 
\begin{eqnarray}
&& \lim_{\tau \to - \infty} \, \lambda_{\mathrm{ph}}(\lambda,\tau) \to 0, \qquad {\mathrm{whenever}} \qquad \dot{a} >0, \qquad \ddot{a} >0, 
\label{onea}\\
&& \lim_{\tau \to - \infty} \, \lambda_{\mathrm{ph}}(\lambda, \tau) \to \infty, \qquad {\mathrm{whenever}} \qquad \dot{a} <0, \qquad \ddot{a} <0,
\label{oneb}
\end{eqnarray}
where the overdot denotes a derivation with respect to the cosmic time coordinate $t$ which is related to $\tau$ as $a(\tau) \, d\tau = d\,t$. In the context of conventional inflationary scenarios the limit (\ref{onea}) applies since the background is characterized by a phase of accelerated expansion (i.e. $\dot{a} > 0$ and $\ddot{a} > 0$) with mildly decreasing curvature. As an example, for a power-law inflationary expansion [i.e. $a(t) \sim (t/t_{1})^{\alpha}$ with $\alpha > 1$ and $t > t_{1}$] we have $a(\tau) \sim (- \tau/\tau_{1})^{-\beta}$ with $\beta = \alpha/(\alpha -1)$ so that the condition (\ref{onea}) is verified. Similarly in the instance of a power-law contraction $a(t) \sim (- t/t_{1})^{\gamma}$ (with $0<\gamma<1$ and $t < - t_{1}$) the limit of Eq. (\ref{oneb}) easily follows. The evolution equations for the scalar and tensor modes of the geometry cannot be valid for arbitrarily short wavelengths as it seems instead implied by Eq. (\ref{onea}): this is the same kind of aporia appearing in the context of Hawking radiation \cite{jac1,unr1,deser1,deser2,park1}. 

In the context of the conventional inflationary models the limit $\tau \to -\infty$ should be handled with some care since an ever expanding inflationary evolution is not past geodesically complete \cite{FFF4a}. With this caveat, in terms of the physical frequencies the two limits of Eqs. (\ref{onea}) and  (\ref{oneb}) are interchanged:
\begin{eqnarray}
&& \lim_{\tau \to - \infty} \, \omega(k,\tau) \to \infty, \qquad {\mathrm{whenever}} \qquad \dot{a} >0, \qquad \ddot{a} >0, 
\label{twoa}\\
&& \lim_{\tau \to - \infty} \, \omega(k,\tau) \to 0, \qquad {\mathrm{whenever}} \qquad \dot{a} <0, \qquad \ddot{a} <0, 
\label{twob}
\end{eqnarray}
where we used the notation $\omega(k, \tau) = k/a(\tau)$ together with $k = 2\pi/\lambda$. 
Thus the physical frequencies may become easily super-Planckian (or trans-Planckian as some like to say) in the case of conventional inflationary models and sub-Planckian for bouncing scenarios based on a stage of accelerated contraction. All in all  
 the same class of problems originally pointed out in the context of black hole evaporation arise in conventional inflationary scenarios but, apparently, not in the case of bouncing scenarios, at least in the limit $\tau \to - \infty$ \cite{FFF1,FFF2}. To be fair bouncing models are not totally immune from these potential issues as sometimes affirmed. In fact Eqs. (\ref{onea}) and (\ref{oneb}) should probably be 
 complemented by two supplementary limits:
\begin{eqnarray}
&& \lim_{\tau \to 0^{-}} \, \lambda_{\mathrm{ph}}(\lambda,\tau) \to \infty, \qquad {\mathrm{whenever}} \qquad \dot{a} >0, \qquad \ddot{a} >0, 
\label{tena}\\
&& \lim_{\tau \to 0^{-}} \, \lambda_{\mathrm{ph}}(\lambda, \tau) \to 0, \qquad {\mathrm{whenever}} \qquad \dot{a} <0, \qquad \ddot{a} <0.
\label{tenb}
\end{eqnarray}
Equations (\ref{tena}) and (\ref{tenb}) together with Eqs. (\ref{onea})--(\ref{oneb}) suggest that in the case of accelerated expansion 
the physical wavelengths can become smaller than the Planck length close to the onset of inflation but the same issue occurs
at the end of a phase of accelerated contraction.  In practice 
this possibility is also related to the so-called gradient instability stipulating that 
either the scalar or the tensor modes of the geometry may inherit an imaginary sound speed (see for 
instance \cite{FFF5a,FFF5b}). This phenomenon normally occurs when the contracting evolution turns 
 into the expanding stage: since the transition  involves either higher derivatives in the matter fields or 
 higher-order curvature corrections an effective sound speeds for the scalar and tensor modes develops.

The implications of Eqs. (\ref{onea})--(\ref{oneb}) and Eqs. (\ref{twoa})--(\ref{twob}) depend on the initial Cauchy hypersurface that can be assigned in two complementary ways.  According to the first strategy the various modes
 are initialized on a given space-like hypersurface {\em at the same initial conformal time $\tau_{i}$}. This approach is physically meaningful for classical fluctuations that are assigned, once forever, at the onset of inflation (and subsequently ironed provided the inflationary stage is sufficiently long \cite{FFF4b,FFF4c}). According to the second strategy the different modes of the field are assigned {\em on different space-like  hypersurfaces}: in this context  we will have that $\tau_{i} = \tau_{i}(k)$ so that the initial normalization depends on the comoving wavenumber. This second perspective cannot be applied to classical fluctuations but it is physically meaningful for the quantum inhomogeneities that keep on reappearing all along an initial 
inflationary (or bouncing) stage. These two possibilities will now be separately scrutinized and swiftly compared in the light of the conditions of Eqs. (\ref{onea})--(\ref{oneb}) and (\ref{twoa})--(\ref{twob}).  

Let us therefore consider a stage of accelerated expansion and assume that {\em all the $k$-modes are assigned at the same time $\tau_{i}$}. The only way to prevent the presence of super-Planckian frequencies is to demand that $\omega(k, \tau_{i}) \leq M$ for all the different $k$-modes at the same time $\tau_{i}$; more precisely 
\begin{equation}
\omega(k,\tau_{i}) = \frac{k}{a(\tau_{i})} \leq M < M_{P}, \qquad H\, a = - \frac{1}{(1 - \epsilon) \tau},\qquad \epsilon = - \dot{H}/H^2,
\label{three}
\end{equation}
where $M$ is a given physical scale that does not exceed, by construction, the Planck (or string) energy scale; $\epsilon$ is the standard slow-roll parameter. The condition (\ref{three}) is valid for all the modes of the spectrum 
provided  it is verified for the maximally amplified wavenumber $k_{max}$.  For the sake of concreteness we can consider the case of the curvature inhomogeneities and the tensor modes induced by a single scalar field $\varphi$ where $z = a \varphi^{\prime}/{\mathcal H}$ and ${\mathcal H} = a^{\prime}/a \equiv a\, H$; in this situation the maximal wavenumbers for the scalar and tensor modes (i.e. $k_{s\,\,max}$ and $ k_{t\,\,max}$) are determined from the following pair of conditions:
\begin{equation}
k_{s\,\,max}^2 = \frac{z^{\prime\prime}}{z} \biggl|_{\tau = \tau_{f}}, \qquad k_{t\,\,max}^2 = \frac{a^{\prime\prime}}{a} \biggl|_{\tau = \tau_{f}},\qquad \epsilon(\tau_{f}) \simeq \eta(\tau_{f}) = {\mathcal O}(1),
\label{four}
\end{equation}
where the two expressions at the right-hand side of both equations are evaluated at the time $\tau_{f}$ that conventionally 
defines the final stage of the inflationary expansion; around $\tau_{f}$ the slow-roll parameters (i.e. $\epsilon$ and $\eta$) 
are all ${\mathcal O}(1)$.  Neglecting the minor differences induced in Eq. (\ref{four}) by the slow-roll parameters 
we will have that\footnote{The slow-roll parameters are defined, in what follows, as $\epsilon = - \dot{H}/H^2$ (already introduced in Eq. (\ref{three})),  $\eta = \ddot{\varphi}/(H \, \dot{\varphi})$ and $\overline{\eta} = (\epsilon - \eta) \equiv \overline{M}_{P}^2 (W_{\,\varphi\varphi}/W)$, where $\overline{M}_{P} = M_{P}/\sqrt{8 \pi}$ and $W$ denotes the inflaton potential. } 
\begin{equation}
k_{s\, max} \simeq k_{t\, max} = a_{f} H_{f} \biggl[ 1 + {\mathcal O}(\epsilon_{f})  + {\mathcal O}(\eta_{f})\biggr],
\label{five}
\end{equation}
so that the common values of $k_{s\, max}$ and $k_{t\, max}$ can be approximately estimated by $k_{max} \simeq a_{f}\, H_{f} \simeq a_{f} \, H$
where $H$ denotes the Hubble rate during inflation. Equations (\ref{four}) and (\ref{five}) imply that the condition of Eq. (\ref{three}), if applied 
to $k_{max}$, leads to the following chain of inequalities:
\begin{equation}
\frac{k_{max}}{a(\tau_{i})} \leq M\,\, \Rightarrow H\, \biggl(\frac{a_{f}}{a_{i}}\biggr) \leq M\,\,\,\,\,\, \Longrightarrow \,\,\,e^{N} \leq \frac{M}{H},
\label{six}
\end{equation}
where $N = \ln{(a_{f}/a_{i})}$ denotes throughout the total number of inflationary $e$-folds. The last inequality of Eq. (\ref{six}) demands an upper limit on the total number of inflationary $e$-folds.  For actual estimates we shall always choose 
$M= {\mathcal O}(M_{P})$ (even if we shall insist later on that $M< M_{P}$ ); 
in this case Eq. (\ref{six}) suggests that $ N < - \ln{(H/M_{P})}$. Assuming the temperature correlations from the Cosmic Microwave Background are due to the curvature 
inhomogeneities amplified during inflation we have that the inflationary curvature scale is given by $H/M_{P} = \sqrt{ \pi\, r_{T}\, {\mathcal A}_{{\mathcal R}}}$ where ${\mathcal A}_{{\mathcal R}} = {\mathcal O}(2.4)\times 10^{-9}$ is the amplitude of the scalar modes and  $r_{T}$ is the tensor to scalar ratio\footnote{We shall be assuming throughout the validity of the consistency relations stipulating that $16 \epsilon = r_{T} = - 8 n_{T}$ where $n_{T}$ is the tensor spectral index \cite{FFF5}. If the consistency relations are not assumed some of the constraints discussed below could be probably relaxed.}.  Equation (\ref{six}) implies that $N$ can be at most ${\mathcal O}(14)$ supposing $r_{T} = {\mathcal O}(0.01)$. 
This result has been interpreted in Refs. \cite{FFF1,FFF2} as a prohibitive condition forbidding, in practice, the existence of sufficiently long inflationary phases. The results of Eqs. (\ref{three}), (\ref{four})--(\ref{five}) and (\ref{six}) demonstrate  that these severe constraints follow  from the assumption that all the different wavelengths of the scalar and tensor mode of the geometry are assigned on the same Cauchy hypersurface.  It might be useful to remark that  in Refs. \cite{FFF1,FFF2} the condition (\ref{six}) has been simply assumed as a conjecture with the purpose of deriving a certain number of restrictions on the properties of inflationary 
potentials. In the present approach the condition (\ref{six}) is not a conjecture but rather a result that only follows by assigning all the modes of the quantum fields at the same initial time $\tau_{i}$. Even if the two viewpoints are formally different they are in practice 
complementary and this is the reason why it seems useful to investigate how the condition (\ref{six}) can be evaded or superseded.

For a successful inflationary evolution we must fit the event horizon at the onset of inflation within 
the present size of the Hubble radius. Since the typical size of the event horizon at $\tau_{i}$ is ${\mathcal O}(H^{-1})$ we should require that $(a_{0}/a_{i}) = (H/H_{0})$ where $a_{0}$ and $H_{0}$ denote the scale factor and the Hubble rate at the present epoch while $a_{i}$ is the scale factor at the onset of inflation\footnote{We are assuming here that inflation starts on a time scale  ${\mathcal O}(\tau_{i})$.}. This requirement can be made more explicit:\begin{equation}
\frac{H_{0}}{H}= \sqrt{2 \,\Omega_{R0}}\, e^{ - 2 N}, \qquad e^{N} \simeq \frac{M_{P}}{H},
\label{sixa}
\end{equation}
where $\Omega_{R0}$ is the critical fraction of the energy density attributed to massless 
species in the concordance paradigm (i.e. $h_{0}^2 \, \Omega_{R0} = 4.15 \times 10^{-5}$).
The second relation in Eq. (\ref{sixa}) gives the maximum number of $e$-folds compatible 
with Eq. (\ref{six}); if  the two expressions in Eq. (\ref{sixa}) are combined and solved with 
respect to $N$ we get a critical number of $e$-folds, i.e. $N_{c} = {\mathcal O}(45.3)$.
Thus we must have $r_{T} < 16/( \pi \, {\mathcal A}_{{\mathcal R}}) e^{ - 2 N_{c}}$, i.e. 
$r_{T} < 9.5 \times 10^{-31}$. According to this estimate any tensor mode potentially detected 
through the $B$-mode autocorrelations should not be attributed to a conventional inflationary stage (see
e.g. \cite{FFF5}). 

The prohibitive limits explored in the two previous paragraphs ultimately follow from Eqs. (\ref{three}) and (\ref{five}) 
suggesting that all the different $k$-modes should be assigned on the same space-like hypersurface. If this strategy is adopted the maximal amplified frequency exceeds the Planck scale at $\tau_{i}$ for any sound duration of inflation given the current observational bounds on the tensor to scalar ratio. According to the viewpoint conveyed here, the approach leading to Eqs. (\ref{six}) and (\ref{sixa}) is reasonable in the case of classical fluctuations. Conversely there is no compelling reason why all the wavelengths of the scalar and tensor modes of the geometry should be assigned on the same space-like hypersurface when the inhomogeneities are generated quantum mechanically. 
This observation is not new and has be been discussed, with different techniques and motivations, in Refs.  \cite{kan1,dan,mgio1,mgio2} and it responds to the logic of effective theories \cite{FFF4e}.  The different $k$-modes are then assigned as soon as the corresponding physical frequencies ``cross''  the scale $M$:
\begin{equation}
\omega(k, \tau_{i}(k)) = \frac{k}{a[\tau_{i}(k)]} = M < M_{P}.
 \label{seven}
 \end{equation}
 The condition (\ref{seven}) defines, in practice, what we could call Planckian hypersurface even if different 
 nomenclatures exist in the literature and they reflect slightly different physical interpretations. It is important to appreciate that while in Eq. (\ref{three}) $\tau_{i}$ is the same for different $k$-modes, in Eq. (\ref{seven})
$\tau_{i} = \tau_{i}(k)$ and the specific $k$-dependence is ultimately dictated by the 
dynamics of the background. For a phase of accelerated expansion we have\footnote{In Eq. (\ref{eight}) 
we introduced, for later convenience, the notation $x_{i}(k)=  k\, \tau_{i}(k)$.}: 
\begin{equation}
\tau_{i}(k) = - \tau_{f} \biggl(\frac{M}{k}\biggr)^{1 - \epsilon}\,\, \Rightarrow \,\,x_{i}(k)=  k\, \tau_{i}(k) = (- k \,\tau_{f})^{\epsilon} \biggl(\frac{M}{H} \biggr)^{1 - \epsilon} \gg 1.
\label{eight}
\end{equation}
Equation (\ref{eight}) requires that long wavelengths (i.e. small $k$-modes) are normalized ealrier 
while short wavelengths (i.e. large $k$-modes) are normalized later. If 
$\tau_{*}$ conventionally marks the onset of inflation, the smallest wavenumber of the 
spectrum (i.e. $k_{*} \simeq \tau_{*}^{-1}$)
will be assigned at $|\tau_{i}(k_{*})| \gg |\tau_{i}(k_{max})|$ where $\tau_{i}(k_{max})$ 
denotes instead the time at which the largest mode of the spectrum is initialized 
(see Eqs. (\ref{four})--(\ref{five}) for the definition of $k_{max}$). Note that in the limit 
$\epsilon \to 0$ (which is mathematically convenient for order of magnitude estimates) 
we have that $|\tau_{i}(k_{*})| \simeq e^{N} |\tau_{i}(k_{max})|$ where $N$ is the total number 
of $e$-folds. Let us finally consider a phase of accelerated contraction where the scale factor 
can be parametrized, in cosmic time, as $a(t) = (-t/t_{1})^{\gamma}$ with $0<\gamma< 1$ and $t < - t_{1}$.
The condition (\ref{seven}) implies, in this case,
\begin{equation}
\tau_{i}(k) = - \tau_{1} \, \biggl(\frac{k}{M}\biggr)^{ -1 + 1/\gamma}\,\, \Rightarrow \,\,x_{i}(k)=  k\,\tau_{i}(k) = (- k \tau_{1})^{1/\gamma} \, \biggl(\frac{H_{1}}{M}\biggr)^{-1 + 1/\gamma},
\label{nine}
\end{equation} 
where $H_{1} < M_{P}$ represents the maximal curvature scale for $ t = {\mathcal O}(t_{1})$ and $|k \tau_{1} |< 1$.
The condition (\ref{nine}) does not guarantee that the physical frequencies will remain smaller than the Planck mass towards the end of the bouncing phase, as already stressed by Eqs. (\ref{tena}) and (\ref{tenb}) in the case of the corresponding wavelengths.

If the different $k$-modes are assigned as in Eq. (\ref{eight}) the constraint of Eq. (\ref{six}) 
does not arise. More specifically, the total duration of the inflationary phase is not constrained since the maximal 
frequency coincides with $M$ and it is always smaller than $M_{P}$ by construction:
\begin{equation}
\frac{k_{max}}{a[\tau_{i}(k_{max})]} = M \leq M_{P}.
\label{ninea}
\end{equation}
The approach based on Eq. (\ref{ninea}) leads to the correct form of 
the scalar and tensor power spectra together with a series of oscillating corrections controlled,
 for each $k$-mode, by the dimensionless parameter $x_{i}(k)$ introduced in Eqs. (\ref{eight})--(\ref{nine}).  
 The oscillating contributions however, do not solely depend on Eqs. (\ref{seven}) and (\ref{eight}) but also on the way the initial vacuum state is defined, i.e. on which Hamiltonian is minimized  \cite{mgio1,mgio2} at $\tau_{i}(k)$. Since the problem depends upon time, there is always the possibility of performing a time-dependent canonical transformation that changes the explicit form of the Hamiltonian without affecting  the classical evolution. The different Hamiltonians will be minimized by different vacua at $\tau_{i}(k)$ and this will ultimately lead to the different corrections of the corresponding power spectra.

To illustrate the different forms of the power spectra following from the minimization of the various Hamiltonians on the Planckian hypersurfaces it is practical to consider the scalar modes of the geometry:
\begin{equation}
H^{(1)}(\tau) = \frac{1}{2} \int \, d^{3} x \biggl[ \frac{\Pi^2}{z^2} \,\, +\,\, z^2 \, \partial_{k} {\mathcal R}\, \partial^{k} {\mathcal R} \biggr], 
\qquad \Pi = z^2 \,\,\partial_{\tau} {\mathcal R}.
\label{twelve}
\end{equation}
where ${\mathcal R}$ denotes the curvature perturbation on comoving orthogonal hypersurfaces and $\Pi$ is 
the conjugate momentum\footnote{ An analog form of the Hamiltonian can be discussed in the case of the tensor modes 
but, for the sake of conciseness, we shall limit the attention to the scalar case which is also the one 
more relevant for the observations of the temperature and polarization anisotropies.}.
If the classical fields are promoted to the status of quantum operators and subsequently represented 
in Fourier space, the Hamiltonian operator corresponding to Eq. (\ref{twelve}) is
\begin{equation}
\hat{H}^{(1)}(\tau) = \frac{1}{2} \int \, d^{3} k \biggl[ \frac{1}{z^2} \, \hat{\Pi}^{\dagger}_{\vec{k}} \hat{\Pi}_{\vec{k}} 
+ k^2 \, z^2 \, \hat{{\mathcal R}}_{\vec{k}}^{\dagger} \, \hat{{\mathcal R}}_{\vec{k}}\biggr],
\label{thirteen}
\end{equation}
where the Fourier amplitudes of the quantum fields and of the canonical momenta are defined by: 
\begin{equation}
\hat{{\mathcal R}}(\vec{x}, \tau) = \frac{1}{(2\pi)^{3/2}} \int d^{3} k \, \hat{{\mathcal R}}_{\vec{k}} \, e^{ - i \vec{k} \cdot \vec{x}}, \qquad \hat{\Pi}(\vec{x}, \tau) = \frac{1}{(2\pi)^{3/2}} \int d^{3} k \, \hat{\Pi}_{\vec{k}} \, e^{ - i \vec{k} \cdot \vec{x}}.
\label{fourteen}
\end{equation}
The Hermiticity of the fields and of the momenta in real space obviously implies that $\hat{{\mathcal R}}_{\vec{k}}^{\,\, \dagger} = \hat{{\mathcal R}}_{-\vec{k}}$ and $\hat{ \Pi}_{\vec{k}}^{\,\, \dagger} = \hat{\Pi}_{-\vec{k}}$. The Hamiltonian (\ref{thirteen}) can be diagonalized at the initial time $\tau_{i}(k)$ in terms of the operator $\hat{Q}_{\vec{k}}$ and $\hat{Q}_{-\vec{k}}$ defined as:
\begin{equation}
\hat{Q}_{\vec{k}} = \frac{1}{\sqrt{2 k}} \biggl\{\frac{\hat{\Pi}_{\vec{k}}[\tau_{i}(k)]}{z_{i}(k)} -
i \, z_{i}(k) \, k \, \hat{{\mathcal R}}_{\vec{k}}[\tau_{i}(k)]\biggr\},
\label{sixteen}
\end{equation}
where $z_{i}(k) = z[\tau_{i}(k)]$; the canonical commutation relations between conjugate field operators demand from Eq. (\ref{sixteen}) that $[\hat{Q}_{\vec{k}}, \hat{Q}_{\vec{p}}^{\dagger} ] = \delta^{(3)}(\vec{k} - \vec{p})$. The diagonal form of the Hamiltonian (\ref{thirteen}) is finally given by: 
\begin{equation}
\hat{H}(\tau_{i}) =\frac{1}{4}\int \,d^{3} k \, k\,\biggl[ \hat{Q}^{\dagger}_{\vec{k}} \hat{Q}_{\vec{k}} 
+ \hat{Q}_{\vec{k}} \hat{Q}^{\dagger}_{\vec{k}} +\hat{Q}^{\dagger}_{-\vec{k}} \hat{Q}_{-\vec{k}} 
+ \hat{Q}_{-\vec{k}} \hat{Q}^{\dagger}_{-\vec{k}} \biggr].
\label{seventeen}
\end{equation}
The state minimizing the Hamiltonian (\ref{seventeen}) at $\tau_{i}(k)$ is defined by $\hat{Q}_{\vec{k}} |0^{(1)}\rangle =0$ and $\hat{Q}_{-\vec{k}} |0^{(1)}\rangle =0$; these two conditions provide the Cauchy data for the evolution equations in the Heisenberg description so that the scalar power spectrum is
\begin{equation}
\langle 0^{(1)},\tau_{i}(k) | \hat{{\mathcal R}}_{\vec{k}}(\tau) \, \hat{\mathcal R}_{\vec{p}}(\tau)| \tau_{i}(k), 0^{(1)} \rangle = \frac{2 \pi^2}{k^3} \,\, {\mathcal P}^{(1)}_{{\mathcal R}}(k,\tau_{i}) \delta^{(3)}(\vec{k} + \vec{p}).
\label{eighteen}
\end{equation}
The explicit form of the power spectrum ${\mathcal P}^{(1)}_{{\mathcal R}}(k,\tau_{i})$ 
contains a leading term (which is the standard scalar power spectrum typical of single-field 
inflationary models) and a series of corrections in the inverse of $k \tau_{i}(k)$. More specifically we will have that 
\begin{eqnarray}
{\mathcal P}^{(1)}_{{\mathcal R}}(k, \tau_{i}) &=& \overline{P}_{{\mathcal R}}(k) \biggl[ 1 + c^{(s)}_{1}(\epsilon, \overline{\eta}) \frac{\sin{[ 2 x_i + c^{(s)}_{0}(\epsilon,\overline{\eta}) \pi]}}{x_{i}} + {\mathcal O}\biggl(\frac{1}{x_{i}^2}\biggr)\biggr]
\nonumber\\
c^{(s)}_{0}(\epsilon, \overline{\eta}) &=& c^{(s)}_{1}(\epsilon, \overline{\eta}) = \frac{1 + 2 \epsilon - \overline{\eta}}{1 - \epsilon}, \qquad \overline{P}_{{\mathcal R}}(k) = \frac{8}{3 \, M_{P}^4} \biggl(\frac{W}{\epsilon}\biggr)
\label{nineteen}
\end{eqnarray}
where, as before, $W$ denotes the inflaton potential. Since $\epsilon$ and $\overline{\eta}$ are both much smaller 
than one when the largest wavelengths exit the Hubble radius during inflation the first oscillating correction
goes as $x_{i}^{-1} \simeq (H/M)$. 

If we now perform a canonical transformation with the appropriate generating functional, the Hamiltonian  (\ref{twelve}) changes its form even if the evolution remains the same. Let us consider for instance the transformation \cite{mgio1}:
\begin{equation}
\hat{{\mathcal R}} \to \hat{q} = z \, \hat{{\mathcal R}}, \qquad \hat{\Pi} \to \hat{\pi} = \hat{q}^{\prime} -  \frac{z'}{z} \hat{q}.
\label{twentyonea}
\end{equation}
In this case the Hamiltonian (\ref{twelve}) changes its form and the final result will be 
\begin{equation}
H^{(1)}(\tau) \to H^{(2)}(\tau) = \frac{1}{2} \int d^{3} x \biggl[ \pi^2 + 2 \pi \,q + (\partial_{k} q)^2\biggr].
\label{twentyone}
\end{equation}
 The Hamiltonian (\ref{twentyone}) can be minimized by following a procedure similar to the  one examined above. The power spectrum will be given this time 
\begin{equation}
\langle 0^{(2)},\tau_{i}(k) | \hat{{\mathcal R}}_{\vec{k}}(\tau) \, \hat{\mathcal R}_{\vec{p}}(\tau)| \tau_{i}(k), 0^{(2)} \rangle = \frac{2 \pi^2}{k^3} \,\, {\mathcal P}^{(2)}_{{\mathcal R}}(k,\tau_{i}) \delta^{(3)}(\vec{k} + \vec{p}),
\label{twentytwo}
\end{equation}
where $| \tau_{i}(k), 0^{(2)} \rangle$ now denotes the state minimizing the Hamiltonian (\ref{twentyone}) 
and ${\mathcal P}^{(2)}_{{\mathcal R}}(k,\tau_{i})$ will now be given by \cite{mgio1}:
\begin{equation}
{\mathcal P}^{(2)}_{{\mathcal R}}(k, \tau_{i}) = \overline{P}_{{\mathcal R}}(k) \biggl[ 1 + c^{(s)}_{2}(\epsilon, \overline{\eta}) \frac{\cos{[ 2 x_i + c_{0}(\epsilon,\overline{\eta}) \pi]}}{x_{i}^2} + {\mathcal O}\biggl(\frac{1}{x_{i}^3}\biggr)\biggr],
\label{twentythree}
\end{equation}
where, this time, $c^{(s)}_{2}(\epsilon, \overline{\eta}) = - c^{(s)}_{0}(\epsilon, \overline{\eta})/2$. A further Hamiltonian canonically related to the one of Eq. (\ref{twentyone}) is 
\begin{equation}
H^{(3)}(\tau) = \frac{1}{2} \int d^{3} x \biggl[ \widetilde{\pi}^2 +  (\partial_{k} q)^2 - \frac{z''}{z} q^2\biggr], \qquad \widetilde{\pi} = q^{\prime}.
\label{twentyfive}
\end{equation}
The transformations $H^{(1)}(\tau) \to H^{(2)}(\tau)$ and $H^{(2)}(\tau) \to H^{(3)}(\tau)$ are both canonical; for instance 
$H^{(2)}(\tau) \to H^{(3)}(\tau)$ corresponds to a generating functional that depends on the old fields $q$,
on the new momenta $\tilde{\pi}$ and on the background evolution 
\begin{equation}
{\mathcal G}( q, \tilde{\pi}, \tau) = \int d^{3} x \biggl( q \tilde{\pi}
 - \frac{z^{\prime}}{2\,z} q^2\biggr).
\label{twentyfivea}
\end{equation}
By differentiating the generating functional, we obtain 
the relation between the old momenta (i.e. $\pi$) and the new ones, 
as well as a change in the Hamiltonian
\begin{equation} 
\pi \to \widetilde{\pi}  = \pi + \frac{z^{\prime}}{z}\, q, \qquad H^{(2)}( q, \pi, \tau) \to H^{(3)}(q,\tilde{\pi},q) = H^{(2)}(q, \pi,\tau) + \frac{\partial{\mathcal G}}{\partial\tau}.
\label{twentyfiveb}
\end{equation}
Bearing in mind Eqs. (\ref{twentyfivea})  the right-hand side of 
Eq. (\ref{twentyfiveb}) leads exactly to Eq. (\ref{twentyfive}). With similar considerations,
all the Hamiltonians (\ref{twelve})--(\ref{twentyfiveb}) can be related to one another 
by suitable canonical transformations. Denoting by $| \tau_{i}(k), 0^{(3)} \rangle$ the state minimizing 
the quantum version of the Hamiltonian $H^{(3)}(\tau)$, the corresponding power 
spectrum will now be given by:
\begin{eqnarray}
{\mathcal P}^{(3)}_{{\mathcal R}}(k, \tau_{i}) &=& \overline{P}_{{\mathcal R}}(k) \biggl[ 1 + c^{(s)}_{3}(\epsilon, \overline{\eta}) \frac{\sin{[ 2 x_i + c^{(s)}_{0}(\epsilon,\overline{\eta}) \pi]}}{x_{i}^3}+{\mathcal O}\biggl(\frac{1}{x_{i}^4}\biggr)\biggr], 
\nonumber\\
c^{(s)}_{3}(\epsilon, \overline{\eta}) &=& - \frac{( 1 + 2 \epsilon - \overline{\eta})(2 + \epsilon- \overline{\eta})}{2 ( 1 - \epsilon)^2}.
\label{twentysix}
\end{eqnarray}
The comparison of Eqs. (\ref{nineteen}), (\ref{twentythree}) and (\ref{twentysix}) demonstrates that the leading term of the spectrum is the same in the three cases even if the corrections are sharply different. The correction 
to the power spectrum goes as $1/x_i^2$ in the case of (\ref{twentythree}); this figure is much smaller than 
the one appearing in (\ref{nineteen}). For instance assuming  $M \sim M_{\rm P}$ the correction will be ${\mathcal O}(10^{-12})$, i.e. six orders of magnitude smaller than in the case of (\ref{nineteen}). In Eq. (\ref{twentysix}) 
 the correction arising from the initial state goes instead as $1/x_i^3$ and, again, if $M \sim M_{\rm P}$ 
it is ${\cal O}(10^{-18})$, i.e. $12$ orders of magnitude smaller than in the case discussed 
in Eq. (\ref{nineteen}). Similar figures arise in the case of the tensor modes 
but will not be specifically discussed here (see \cite{mgio1,mgio2} for further details on this point). The corrections of the scalar power spectra are correlated 
with the rate at which the pump fields of each Hamiltonian go to zero in the 
limit $\tau\to - \infty$: the faster the free Hamiltonian is recovered in the limit $\tau\to - \infty$ 
the smaller is the correction to the power spectrum. This degree of adiabaticity 
 is also correlated with the backreaction effects of the initial vacuum state 
which are negligible in the case of Eqs. (\ref{twentythree}) and (\ref{twentysix}) but not in the case of 
Eq. (\ref{nineteen}) \cite{mgio1}.  

So far all the discussion has been conducted in the Einstein frame. The same 
conclusions can be however reached  in any conformally related frame and, in particular, in the 
string frame. In the string frame the string mass is constant while the Planck mass depends on the value 
of the dilaton coupling $ e^{\varphi/2}$ according to a relation that can be parametrized at low energies as
$M_{s} = e^{\varphi/2} M_{P}$. Recalling that the relation between the metric tensors in the Einstein and in the string 
frames can be written, in four dimensions, as $g_{\mu\nu}^{(e)}= e^{-\varphi} g_{\mu\nu}^{(s)}$ (with $\varphi_{s} = \varphi_{e} = \varphi$), the connection between the scale factors in the two frames will be given by $a_{s} = e^{\varphi/2} a_{e}$.  Assuming for simplicity $M = M_{P}$ we have, for instance, that the condition 
(\ref{seven}) is the same in both frames:
\begin{equation}
\omega_{e}(k,\tau) = \frac{k}{a_{e}(\tau)} < M_{P}\,\,\Rightarrow \,\,  \omega_{s}(k,\tau) = \frac{k}{a_{s}(\tau)} < M_{s} \qquad M_{s} = e^{\varphi/2} M_{P},
\label{twentyseven}
\end{equation}
where the first equation is the same as Eq. (\ref{seven}) while the second relation is its 
analog in the string frame. The metric fluctuations of the geometry in the two frames are in principle different 
however the curvature perturbations on comoving orthogonal hypersurfaces are the same in the two frames, 
${\mathcal R}_{s} = {\mathcal R}_{e}= {\mathcal R}$; a similar property holds for the tensor modes of the 
geometry \cite{FFF5b}. 

Since the quantum mechanical fluctuations are continuously generated during a primeval inflationary (or bouncing) stage, all the different wavelengths of the scalar and tensor modes of the geometry do not have to be assigned on the same space-like hypersurface. We showed that some prohibitive constraints either on the total number of inflationary $e$-folds or on the tensor to scalar ratio can be evaded if the various wavelengths are assigned as soon as their associated physical frequency is still smaller than the Planck (or string) scale.  According to this strategy the initial Cauchy data for the mode functions effectively depend on the wavenumber so that larger wavelengths start their evolution earlier than those that are comparatively shorter. The same problems in conventional inflationary scenarios also occurs in the context of backgrounds with different kinematical properties (e.g.  contracting stages): 
while in the case of an accelerated expansion the physical wavelengths get potentially
shorter than the Planck length at the onset of inflation, for a phase of accelerated contraction the physical wavelengths suffer the same problem but at the end of the bouncing regime (when the scale factor shrinks and the absolute value of the curvature increases). The oscillating contributions arising in the large-scale power spectra as a result of the normalization on the Planckian hypersurfaces turn out to be arbitrarily small. The present result suggest that the duration of a conventional inflationary phase and the tensor to scalar ratio remain unconstrained if the quantum inhomogeneities are appropriately assigned on the Planckian hypersurfaces.

\section*{Acknowledgements}
The author wishes to thank T. Basaglia, A. Gentil-Beccot, S. Rohr and J. Vigen of the CERN Scientific Information Service for their kind assistance.

\end{document}